\documentclass{INTERSPEECH2023}
\usepackage{stmaryrd}

\interspeechcameraready

\title{Privacy-preserving Representation Learning for Speech Understanding}
\name{Minh Tran and Mohammad Soleymani}
\address{
  Institute for Creative Technologies, University of Southern California, Los Angeles, USA}
\email{\{mtran, soleymani\}@ict.usc.edu}

\begin{document}

\maketitle
 
\begin{abstract}
Existing privacy-preserving speech representation learning methods target a single application domain. In this paper, we present a novel framework to anonymize utterance-level speech embeddings generated by pre-trained encoders and show its effectiveness for a range of speech classification tasks. Specifically, given the representations from a pre-trained encoder, we train a Transformer to estimate the representations for the same utterances spoken by other speakers. During inference, the extracted representations can be converted into different identities to preserve privacy. We compare the results with the voice anonymization baselines from the VoicePrivacy 2022 challenge. We evaluate our framework on speaker identification for privacy and emotion recognition, depression classification, and intent classification for utility. Our method outperforms the baselines on privacy and utility in paralinguistic tasks and achieves comparable performance for intent classification.
\end{abstract}

\section{Introduction}
\label{sec:intro}
Privacy-preserving data processing is an actively researched topic due to its societal significance that motivated the introduction of privacy protection legislation such as the General Data Protection Regulation (GDPR). Since speech data can disclose personally identifiable information such as age \cite{metze2007comparison}, gender \cite{metze2007comparison}, and race \cite{walton1994speaker}, it is important to develop privacy-preserving technologies for speech processing.

\begin{figure}[t]

\begin{minipage}[b]{1.0\linewidth}
  \centering
  \includegraphics[width=\linewidth,trim={200pt 140pt 400pt 100pt},clip]{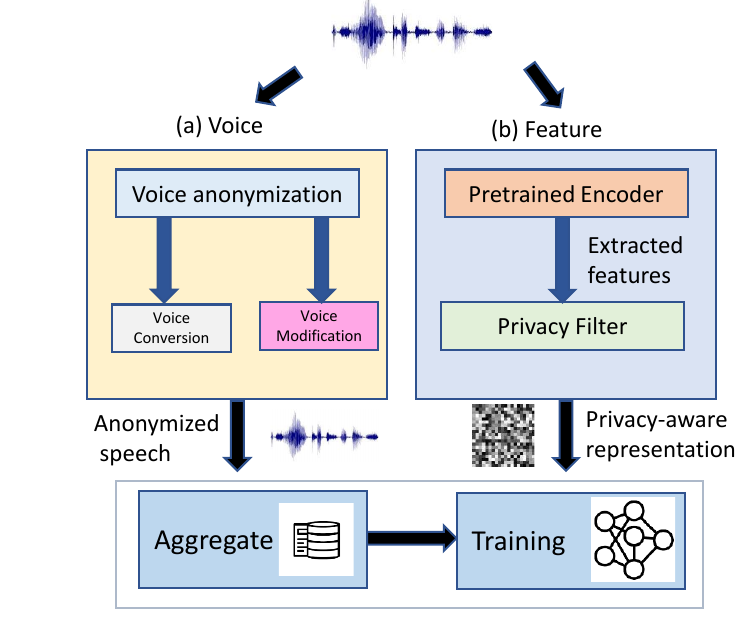}
\end{minipage}
\caption{Existing anonymization approaches focus on voice anonymization that produces speech outputs (a). We explore an alternative direction to sanitize features from pre-trained encoders directly (b).}
\label{fig:compare_voice_ano}
\vspace{-8mm}
\end{figure}



\begin{figure*}[t]
\centering
\includegraphics[width=\linewidth,trim={20pt 330pt 180pt 30pt},clip]{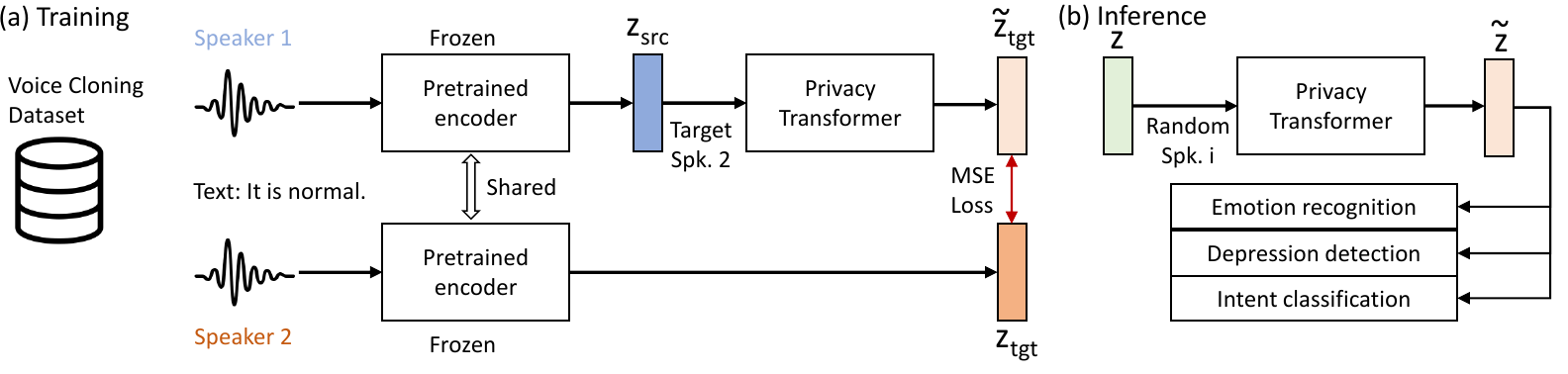}
\caption{An overview of the proposed method. A privacy Transformer is trained to make the features identity-conditioned while preserving their utility for downstream tasks. During inference the Privacy Transformer can be used to sanitize features from identifiable information, thus, preserving privacy.}
\vspace{-2mm}
\label{fig:method_overview}
\end{figure*}

Distributed learning, encryption, and anonymization are the well-known approaches for privacy preservation. While distributed training solutions such as federated learning preserve privacy by only sharing models’ updates, the shared gradients can be vulnerable to privacy attacks \cite{tomashenko2022privacy}. Encryption methods lock the data in such a way that makes it unusable until decrypted \cite{zhang2019encrypted, brasser2018voiceguard} but are computationally expensive. With lower computational requirements, anonymization solutions aim at removing sensitive information from speech signals while preserving everything else. Generally, existing anonymization studies focus on voice anonymization, which alters the voice of the original speaker to hide the speaker’s identity while leaving the remaining information intact \cite{tomashenko2022voiceprivacy, kai2021lightweight, yoo2020speaker, qian2019speech, patino2021speaker}. However, for supervised machine learning purposes, an alternative approach of anonymizing speech at the feature level is under-explored. Because the speech utterance has to be converted to feature representations at some point during the training process, it is possible to locally extract and sanitize the representations and only release the privacy-aware representations instead of a speech utterance. As the features extracted from pre-trained encoders are useful for a wide range of downstream tasks, the anonymized representations are expected to be equally universal.

In this paper, we propose a novel framework to anonymize (remove identifiable information) features extracted using a pre-trained speech encoder (HuBERT \cite{hsu2021hubert}) while preserving other information useful to a wide range of downstream tasks. As an initial step towards feature-level anonymization, we only focus on sanitizing the mean-pooled vector representations from pre-trained encoders, which are commonly used for utterance-level classification tasks. Our \textit{problem formulation} is as follows: Given a vector $z$ extracted for an utterance $u$ via a speech encoder $E$, we want to develop a method $M$ such that $\tilde z=M(z)$ remains useful for different types of downstream tasks while reducing speaker identifiable information.
Our method first trains a Transformer-based \cite{vaswani2017attention} representation learning model that learns to convert the mean-pooled embedding extracted from a source utterance to the embedding of the same utterance spoken by a different speaker, using the VCTK voice cloning database \cite{veaux2017cstr}. During inference, it converts the input embeddings into randomly selected speakers to preserve the speakers' identities. We evaluate our method on a speaker identification task with the VoxCeleb1 dataset \cite{nagrani2017voxceleb} for privacy. For utility, we evaluate our method on emotion recognition (paralinguistic), depression detection (paralinguistic) and intent classification (semantic). Our method is able to outperform the voice anonymization baselines from the VoicePrivacy 2022 Challenge \cite{tomashenko2022voiceprivacy} on the privacy metric and paralinguistic tasks, while achieving comparable performance for intent classification. We further show that our approach is computationally efficient compared to the baselines in terms of runtime and memory usage.

\section{Related Work} \label{sec:related_work}
\noindent \textbf{Self-supervised representation learning}
Self-supervised representation is effective for extracting high-level features from speech signals by learning the underlying structure of the inputs in a self-supervised manner. There are three lines of self-supervised representation learning for speech: generative, contrastive, and predictive. Generative methods generally try to reconstruct masked frames \cite{liu2020mockingjay} or predict future frames \cite{chung2020vector}. Contrastive methods compare pairs of sampled frames with contrastive losses \cite{baevski2020wav2vec, baevski2019vq}. In this work, we focus on sanitizing features produced by HuBERT \cite{hsu2021hubert}, which is an instance of predictive methods that predict the pseudo-labels for masked frames. The discrete pseudo-labels for HuBERT are created by applying K-means clustering on extracted MFCC features. \\
\noindent \textbf{Voice anonymization}
Voice anonymization methods tend to follow two lines of work, namely voice conversion and voice modification. While voice conversion-based solutions deal with changing the x-vector \cite{snyder2018x} for an input utterance via speech synthesis , voice modification-based solutions directly perturb the waveforms using speech processing techniques. Yoo \textit{et al.} \cite{yoo2020speaker} propose using many-to-many voice conversion technique based on Variational Auto-encoders (VAE) to extract and modify speaker identity vectors. Qian \textit{et al.} \cite{qian2019speech} combine sensitive keyword substitution with voice conversion to reduce speaker recognition accuracy. Recently, the VoicePrivacy 2022 Challenge releases its first baseline based on x-vector modification, in combination with the fundamental frequency, bottleneck features, and a speech synthesis component, to produce anonymized speech \cite{tomashenko2022voiceprivacy}. For voice modification, Kai \textit{et al.} \cite{kai2021lightweight} explore various lightweight speech processing methods such as vocal length normalizing, clipping, and resampling to achieve anonymization. The second baseline of the VoicePrivacy 2022 Challenge uses McAdams coefficients \cite{mcadams1984spectral} to shift the pole positions extracted from linear predictive coding analysis for voice anonymization. \\
\noindent \textbf{Privacy-aware representation learning}
Prior work on privacy-preserving representation learning for speech signals generally focus on a single target application. Srivastava \textit{et al.} use adversarial training to learn representations that is useful for automatic speech recognition (ASR) without disclosing speaker identity \cite{srivastava2019privacy}. Jaiswal \textit{et al.} \cite{jaiswal2020privacy} use Gradient Reversal Layer \cite{ganin2016domain} to learn gender-invariant features that perform well for emotion recognition. To the best of our knowledge, there has been no prior work exploring privacy-preserving representations that perform well on multiple applications.
\section{Proposed Method}
\label{sec:method}

Figure \ref{fig:method_overview} shows an overview of our proposed method. At the core of our approach is a Privacy Transformer that converts the identity of the extracted representations from a pre-trained encoder to achieve privacy. In particular, we train our Privacy Transformer with a Voice Cloning dataset that provides speech utterances of different speakers reading a fixed set of sentences. Given a pair of utterances with the same content, our Transformer learns to predict the embedding of the target utterance (speaker 2) given the embedding of the source utterance (speaker 1) and the target speaker. During inference, we use the trained Privacy Transformer to convert extracted embeddings to randomly selected identities to produce privacy-preserving representations that are suitable for different downstream tasks.\\ 
\noindent \textbf{Pretrained encoder.} Although our method is compatible to different speech encoders, we focus our analysis on the HuBERT-base representation learning model \cite{hsu2021hubert} in this work. The architecture of HuBERT consists of a Convolutional-based feature extractor to convert raw waveforms into low-level features, followed by a Transformer encoder \cite{vaswani2017attention} to generate high-level representations. During training, the model first generates pseudo-labels from input audio signals with k-means clustering on extracted MFCC features. The model is then optimized in a self-supervised manner by predicting the masked pseudo-labels.

The Transformer encoder within the HuBERT-base model has $L=12$ layers with an embedding dimension of $d=768$. Because features from different layers of HuBERT contain different information \cite{chang2022distilhubert}, we extract features and temporally mean-pool them from all 12 layers of HuBERT. Hence, each utterance can be represented as a matrix of size $L\times d$.\\
\noindent \textbf{Privacy Transformer.}
The Privacy Transformer P is fed by the mean-pooled features from all 12-layers of HuBERT for the input utterances in addition to the target speakers (discrete IDs). It consists of three main components, namely, embedding layers $P_{emb}$, a Transformer encoder $P_{te}$ and a fully-connected layer $P_{fc}$. We use two trainable embedding layers, namely $P^{spk}_{emb}$ and $P^{layer}_{emb}$, to convert discrete target speakers and layer information (layer 1 to 12) into vectors of size $d_{spk}$ and $d_L$, respectively. The speakers and layer embeddings are then concatenated with the HuBERT embeddings to form inputs of size $bsz \times L \times (d+d_{spk}+d_L)$ for the Transformer encoder. Our Transformer encoder follows closely the original architecture from \cite{vaswani2017attention} that consists of $L_P$ stacks of self-attention layers followed by a feed-forward neural network. However, we do not use the positional encoding as our inputs are permutation-invariant. The goal of the Transformer encoder is to model the relationship between embeddings of different layers (but the same content) to produce high-level layer-aware contextualized embeddings. 
Finally, we use a fully-connected layer to map the output dimension back to $d$. 
Although having a Transformer encoder to post-process the extracted representations might seem computationally expensive at first, P can efficiently process its inputs due to the lack of the temporal dimension, which we later demonstrate via experimental results in the discussion section.\\
\noindent \textbf{Learning scheme.} In each learning step, $N$ pairs of utterances from different speakers with the same content, \textit{i.e.}, $\{(u^i_1, spk^i_1)\}^N_{i=1}$ and $\{(u^i_2, spk^i_2)\}^N_{i=1}$,  are sampled from the Voice Cloning dataset (with speakers pool $S$), and the HuBERT encoder $E$ extracts the feature embeddings for each utterance
\begin{equation}
\begin{aligned}
z_{src}=\{E(u^i_1)\}^N_{i=1} \quad ; \quad  z_{tgt}=\{E(u^i_2)\}^N_{i=1}
\end{aligned}
\end{equation}
 The Privacy Transformer P takes as inputs $z_{src}$ along with the target speakers $spk_{tgt}=\{spk^i_2\}^N_{i=1}$ and generate estimations $\tilde{z}_{tgt}$ for the target representations $z_{tgt}$.
\begin{equation}
\begin{aligned}
\tilde{z}_{tgt}=P(z_{src}|spk_{tgt})=P_{fc}(P_{te}(z_{src} \oplus z_{spk} \oplus z_{layer}))
\end{aligned}
\end{equation}
where $\oplus$ denotes the concatenation operation with 
\begin{equation}
\begin{aligned}
z_{spk}=P^{spk}_{emb}(spk_{tgt})\quad ; \quad z_{layer}=P^{layer}_{emb}(layers)
\end{aligned}
\end{equation}
for $layers=\llbracket 1..L \rrbracket $.
 To achieve this goal, the Privacy Transformer is trained with a mean-squared error loss
 \begin{equation}
 L_P=||\tilde{z}_{tgt}-z_{tgt}||_2     
 \end{equation}
\noindent \textbf{Inference scheme.} During inference, we want to anonymize a set of arbitrary utterances (from unseen speakers), \textit{i.e.}, $\{u^i\}^N_{i=1}$, without any information about the speaker identities. We use $E$ to extract $z=\{E(u^i)\}^N_{i=1}$ features from these utterances. We randomly select target speakers for the input utterances from the speakers pool $S$. 
Here, we allow different layers to have different target speakers because there is no ground-truth target speaker as in the learning process. 
\begin{equation}
\begin{aligned}
spk_{tgt}^{inf}=\{spk_{tgt}^{i}(layers)\}^N_{i=1} \quad ; \quad spk_{tgt}^{i}(layer_j)\stackrel{iid}{\sim} S
\end{aligned}
\end{equation}
The outputs  $\tilde{z}$ from the Privacy Transformer is the sanitized representations produced by our method. 
\begin{equation}
\begin{aligned}
\tilde{z}=P(z_{src}|spk_{tgt}^{inf})
\end{aligned}
\end{equation}
We then use  $\tilde{z}$ for different downstream tasks for privacy and utility evaluations.\\
\noindent \textbf{Choice of the Pretrained encoder and Privacy module.}
Although the experiments presented in this paper are limited to one encoder architecture (HuBERT \cite{hsu2021hubert}) and one Privacy model architecture (Transformer \cite{vaswani2017attention}), the method can be applied to different combinations of pre-trained encoder and privacy module architectures because the roles of the modules are architecture-independent. Specifically, the encoder needs to extract a robust representation of the input speech for a wide range of downstream tasks, and the Privacy module needs to produce speaker-conditioned vector representations, given the extracted features and the target speaker IDs. We chose HuBERT as our feature encoder due to its superior performance and a Transformer as our privacy module due to its ubiquitous usage. To demonstrate the usefulness of our method to Transformer-free encoders, we provide additional experimental results with the APC pre-trained model \cite{chung2019unsupervised} in the supplementary materials.

\section{Experiments}
\label{sec:experiments}
\subsection{Datasets} 
\noindent \textbf{CSTR VCTK} The CSTR VCTK dataset \cite{veaux2017cstr} includes speech data utters by 110 English speakers of various accents. Each speaker reads from a fixed set of about 400 sentences collected from different sources such as a newspaper, a rainbow passage and an elicitation paragraph. For preprocessing, we downsample the 48kHz audio recordings provided by the dataset to 16kHz to be compatible with HuBERT feature extraction.
\begin{table*}[t]
\centering
\begin{tabular}{ccccccccc}
\hline
               & SID Acc. $\downarrow$  & IEM. Acc. & IEM. F1 & MSP. Acc. & MSP. F1 & AVEC Acc. & AVEC F1 & IC Acc. \\ \hline
Original       & 72.79 & 67.01    & 67.97   & 61.29    & 58.10    & 60.65     & 56.12   & 95.86   \\ \hline
Baseline 1     & 14.94 & 60.01    & 61.24   & 52.65    & 46.24   & 53.35         & 52.55       & \textbf{93.37}   \\ 
Baseline 2     & 29.92 & 62.50     & 63.35   & 54.63    & 49.54   & 57.56         & 54.75       & 86.46   \\ 
Baseline 3     & 11.72     & 60.18        & 60.96       & 54.14        & 46.45       & 57.88         & 54.22       & 89.94       \\ \hline
Propose method & \textbf{11.64} & \textbf{63.87}    & \textbf{64.97}   & \textbf{58.17}    & \textbf{54.33}   & \textbf{58.13}     & \textbf{55.98}   & 92.91   \\ \hline
\end{tabular}
\caption{Comparison results between our proposed method and the baselines.}
\label{tab:results}
\vspace{-5mm}
\end{table*}

\begin{table}[t]
\begin{tabular}{ccc}
\hline
                & Runtime (s) & Memory Usage (GB) \\ \hline
HuBERT Extr.    & 439         & 1.13              \\ \hline
Baseline 1      & 2022        & 3.50               \\ 
Baseline 2      & 494         & 0.18              \\ 
Baseline 3      & 457         & 1.13             \\ \hline
Proposed method & 518         & 1.71              \\ \hline
\end{tabular}
\caption{Efficiency comparison between our proposed method and the baselines.}
\label{tab:effiency}
\vspace{-5mm}
\end{table}
\noindent \textbf{VoxCeleb1} We use the VoxCeleb1 dataset \cite{nagrani2017voxceleb} to report the speaker identification (SID) performance, our privacy metric.
The dataset contains over 145K utterances from 1541 speakers, collected from more than 21K YouTube videos. We follow the standard train/test splits provided by the dataset with 10\% of the train set randomly selected for validation.

It is important to note that prior studies generally use (semi-informed) speaker verification models to demonstrate robustness to attacks \cite{tomashenko2020voiceprivacy, tomashenko2022voiceprivacy}. However, it is rather difficult for us to construct a robust speaker verification model given that our inputs are 1D vector representations of speech. Therefore, we report the SID performance in this work, as a speaker verification-based attack is more appropriate for future work on anonymizing the entire speech representations with a temporal dimension.\\
\noindent \textbf{IEMOCAP} The IEMOCAP dataset \cite{busso2008iemocap} is a commonly used dataset for multi-modal emotion recognition with approximately 12 hours of recordings from 10 speakers. The dataset is collected in 5 sessions, in which participants are asked to act according to scripts that are designed to invoke emotions. Following prior work \cite{yang2021superb}, We focus on the four basic emotions (happy, sad, angry, and neutral) to avoid severe class imbalance, which contains 5,536 utterances. \\
\noindent \textbf{MSP-IMPROV} The MSP-IMPROV dataset \cite{busso2016msp} is an acted audio-visual emotional dataset that contains acted and improvised emotional dyadic interactions from 12 speakers over 6 sessions. In total, the dataset contains around 8.5K utterances (over 9 hours). Similar to IEMOCAP, we drop the unbalanced emotion classes and focus on the four basic emotion, which reduce the number of utterances to 7,735.\\
\noindent \textbf{Extended-DAIC} The E-DAIC dataset \cite{devault2014simsensei} contains 219 video recordings of clinical interviews probing for symptoms of psychological distress. It was used as part of the 2019 Audio-Visual Emotion Challenge Workshop (AVEC2019) \cite{ringeval2019avec} in the Detecting Depression sub-challenge. Each interview, ranging from 15 to 25 minutes, is rated according to the eight-item Patient Health Questionaire (PHQ-8) along with a binary label for depression conditions. We use the provided transcripts with the dataset to extract 18,846 utterances, from which around 33\% are labeled as having depression.\\
\noindent \textbf{Fluent Speech Commands} The Fluent Speech Commands dataset is a widely used dataset for the semantic task of intent classification. It contains around 30,000 utterances from 97 speakers with 31 unique intents used for controlling smart-home appliances and virtual assistants.
\vspace{-2mm}
\subsection{Baselines} 
We use the two baselines provided by the VoicePrivacy 2022 Challenge \cite{tomashenko2022privacy}. \textit{Baseline 1} is based on the voice anonymization architecture proposed in \cite{fangspeaker}. The method first extracts relevant features from the input speech signal, including the x-vector \cite{snyder2018x}, the fundamental frequency, and bottleneck features. The extracted x-vector is then anonymized with an external pool of x-vectors, and the anonymized speech is synthesized with a speech synthesis model based on the anonymized x-vector. We use a variant of the first baseline, in which the bottleneck features are replaced with representations produced by a finetuned wav2vec 2.0 model \cite{baevski2020wav2vec} and the speech synthesis module is HifiGAN-based \cite{kong2020hifi}. \textit{Baseline 2} is based on McAdams transformation \cite{patino2021speaker} with a uniformly sampled McAadam coefficient. We add \textit{Baseline 3} based on Differential Privacy \cite{dwork2008differential} with the mean-pooled embeddings. In particular, the embeddings extracted from HuBERT are clipped to $[-1,1]$ and Laplacian noise $\sim Lap(\frac{2}{\epsilon})$ is added independently to all dimensions.

\subsection{Implementation Details} For the Privacy Transformer, we set the size of the speaker embeddings $d_{spk}=256$ and size of the layer embedding $d_L=128$. Our Privacy Transformer contains of $L_P=5$ stacks of Transformer encoder layers with $h=8$ self-attention heads and an intermediate size of $4608$. We train our Privacy Transformer with an SGD optimizer with a learning rate of $0.001$ for $50$ epochs. During the evaluation, our classification model consists of a featurizer (12 learnable scalars) to fuse information from all 12 layers of the extracted features from HuBERT, followed by 2 fully-connected layers of sizes $\{256,128\}$. For a fair comparison, all methods are trained using the Adam optimizer with a learning rate of $1e^{-3}$ for 50 epochs with early stopping. For baselines 1 and 2, we extract HuBERT features after the original utterances are converted to privacy-preserving utterances. 

\section{Results and Discussion}
\label{sec:results}
Table \ref{tab:results} shows the performance comparisons between our method with the baselines. Since \textit{Baseline 3} allows flexibility in terms of privacy and utility trade-offs, we adjust the amount of added noise ($\epsilon=15$) such that the Speaker Identification accuracy is similar to our method. We also report the original performance of HuBERT (without any privacy-preserving processing) in the first row of Table \ref{tab:results}. Surprisingly, \textit{Baseline 3} achieves a very competitive performance on both privacy and utility metrics compared to the voice anonymization approaches despite involving minimal post-processing. Our method achieves similar speaker identification accuracy compared to the most secure baseline (11.64\% vs. 11.72\%) while outperforming all of the baselines on paralinguistic tasks. Specifically, for emotion recognition, our method achieves F1-scores of 64.97\% (IEMOCAP) and 54.33\% (MSP-IMPROV) compared to 63.35\% and 49.54\% for \textit{Baseline 2}. For depression detection, our approach also outperforms \textit{Baseline 2} with a margin of 1.23\% on the F1 score. For intent classification (a semantics task), the proposed approach achieves comparable classification accuracy with the best-performing baseline (92.91\% vs. 93.37\%). The results indicate feature-level anonymization's potential compared to voice anonymization for supervised ML tasks.

Because anonymization techniques are developed to run locally with limited computation, we compare our method with the baselines in terms of runtime and memory usage. In particular, we select a fixed set of 500 utterances from the VoxCeleb1 dataset \cite{nagrani2017voxceleb} and perform anonymization on four CPUs with a batch size of 1. Table \ref{tab:effiency} shows the efficiency comparisons between our method and the baselines. We also report the computational resources for independently extracting HuBERT features from the selected utterances in the first row. Adding the Privacy Transformer post-processing increases the computation costs by 18\% in runtime and 51\% in memory usage compared to \textit{HuBERT Extr} (extract HuBERT features from speech without any post-processing). Our method achieves superior efficiency performance compared to \textit{Baseline 1} (the more secured voice anonymization baseline) and comparable runtime to \textit{Baseline 2}. Although \textit{Baseline 2} achieves superior efficiency in terms of memory usage, the method is the weakest privacy-preserving baseline and also suffers from poor performance on the semantic task. It is worth noting that the performance of our method is highly dependent on the efficiency of the speech encoders, and hence, there is still room to further boost the efficiency with light encoders such as the Distill-HuBERT \cite{chang2022distilhubert}.

\section{Conclusion}
\label{sec:conclusion}
We present a novel framework to anonymize mean-pooled vector speech embeddings extracted from pre-trained HuBERT. We train a Privacy Transformer to convert the extracted embeddings into different identities while preserving the content using a voice cloning dataset. We compare our method with the VoicePrivacy 2022 Challenge anonymization baseline methods on the following tasks: speaker identification (privacy), emotion recognition (utility), depression recognition (utility), and intent classification (utility). Our method achieves superior privacy preservation performance and outperforms the baselines on paralinguistic tasks. We further show that our method is computationally efficient, compared to the baselines.

\section{Acknowledgement}
Research was sponsored by the Army Research Office and was accomplished under Cooperative Agreement Number W911NF-20-2-0053. The views and conclusions contained in this document are those of the authors and should not be interpreted as representing the official policies, either expressed or implied, of the Army Research Office or the U.S. Government. The U.S. Government is authorized to reproduce and distribute reprints for Government purposes notwithstanding any copyright notation herein.

\bibliographystyle{IEEEtran}
\bibliography{mybib}

\end{document}


\maketitle

\begin{table*}[t]
\centering
\begin{tabular}{ccccccc}
\hline
               & SID Acc. $\downarrow$  & IEM. Acc. & IEM. F1 & MSP. Acc. & MSP. F1 & IC Acc. \\ \hline
Original       & 42.31 & 64.15    & 61.47   & 51.24    & 45.07  & 71.46    \\ \hline
Baseline 1     & 9.25 & 55.16    & 54.69   & 46.71    & 37.29  & 65.40    \\ 
Baseline 2     & 22.46 & 59.28     & 57.96   & 48.83    & 41.51   & 62.16  \\ 
Baseline 3     & 8.36     & 58.14        & 57.28       & 46.32        & 40.09  & \textbf{66.29}           \\ \hline
Propose method & \textbf{7.84} & \textbf{60.21}    & \textbf{58.65}   & \textbf{49.47}    & \textbf{42.31} & 64.48 \\ \hline
\end{tabular}
\caption{Comparison results between our proposed method and the baselines using features from the APC encoder \cite{chung2019unsupervised}.}
\label{tab:results}
\vspace{-5mm}
\end{table*}

 \section{Additional Results}
 We provide additional experimental results for the  Autoregressive Predictive Coding (APC) pre-trained model \cite{chung2019unsupervised} to demonstrate the applicability of our proposed method on Transformer-free encoders. The backbone of the encoder is a deep LSTM model that learns in an unsupervised and autoregressive manner to predict future information given the current and past frames. We use similar experimental settings as described in the paper. The degraded performance across tasks is due to the weaker representation of APC compared to HuBERT \cite{hsu2021hubert}. We do not report results for the depression detection task since the performance is close to random guessing even for the features extracted from the original utterances.

We can see that the conclusions mostly hold despite using a different speech encoder architecture. Our method maintains its superior performance across two emotion recognition datasets while producing privacy-preserving speech embeddings that achieve the lowest speaker identification accuracy. For the semantic task of intent classification, the method achieves stronger performance compared to the two voice anonymization baselines from the VoicePrivacy Challenge 2022 \cite{tomashenko2022voiceprivacy} but performs slightly worse ($1.81\%$ lower accuracy) than \textit{Baseline 3}, which adds Laplacian noise directly to the speech embeddings. Nevertheless, the experimental results demonstrate that the proposed Privacy Transformer remains effective in producing useful privacy-preserving embeddings across different tasks for Transformer-free encoders.

\bibliographystyle{IEEEtran}
\bibliography{mybib}